# Versatile spaceborne photonics with chalcogenide phase-change materials

Hyun Jung Kim[1*], Matthew Julian[2], Calum Williams[3], David Bombara[4], Juejun Hu[5,6],
Tian Gu[5,6], Kiumars Aryana[1], Godfrey Sauti[1], and William Humphreys[1]

[1]*NASA Langley Research Center, Hampton, Virginia, USA*
[2]*Booz Allen Hamilton, Arlington, Virginia USA*
[3]*Department of Physics, University of Cambridge, Cambridge, CB3 0HE, UK*
[4]*John A. Paulson School of Engineering and Applied Sciences, Harvard University, Cambridge, Massachusetts, USA*
[5]*Department of Materials Science & Engineering, Massachusetts Institute of Technology, Cambridge, Massachusetts, USA*
[6]*Materials Research Laboratory, Massachusetts Institute of Technology, Cambridge, Massachusetts, USA*
**\* Corresponding author: hyunjung.kim@nasa.gov**

**Abstract**
Recent growth in space systems has seen increasing capabilities packed into smaller and lighter Earth observation and deep space mission spacecraft. Phase-change materials (PCMs) are nonvolatile, reconfigurable, fast-switching, and have recently shown a high degree of space radiation tolerance, thereby making them an attractive materials platform for spaceborne photonics applications. They promise robust, lightweight, and energy-efficient reconfigurable optical systems whose functions can be dynamically defined on-demand and on orbit to deliver enhanced science or mission support in harsh environments on lean power budgets. This comment aims to discuss the recent advances in rapidly growing PCM research and its potential to transition from conventional terrestrial optoelectronics materials platforms to versatile spaceborne photonic materials platforms for current and next-generation space and science missions. Materials International Space Station Experiment-14 (MISSE-14) mission-flown PCMs outside of the International Space Station (ISS) and key results and NASA examples are highlighted to provide strong evidence of the applicability of spaceborne photonics.

The space sector has witnessed tremendous growth within the past decade from not only government agencies but also entrants from the private sector. This is attributed to: (1) the increased availability and cost reduction of launch platforms, (2) expanded utilization of Low Earth Orbit (LEO) for crewed and uncrewed missions, and (3) the development of plans for lunar missions and ones further into the solar system. As a result, sector growth is expected to accelerate in the coming decades[1]. A host of complementary technological innovations and capabilities are underpinning this growth from reusable rockets to miniaturized satellites. These are enabling applications including planetary exploration missions, remote sensing studies and high-speed satellite broadband. Spaceborne sensing, imaging and communications capabilities are provided through electronic and photonics subsystems. Terrestrially, photonic components are rapidly displacing electronic counterparts to enable next-generation power efficient, high speed and highly versatile technologies, from reconfigurable photonic integrated circuits to tunable optical filters[2]. The technological demands of next-generation space systems are yet more demanding, and require operation in harsh environments while constrained by lean SWaP-C (Size, Weight, Power, and Cost) budgets. To realize subsystems that meet such performance demands requires novel photonic material platforms and innovative multifunctional design schemes.

*Chalcogenide phase-change materials* (PCMs) have been proposed for spaceborne solid-state memory modules because of their nonvolatile, reconfigurable, fast-switching and space radiation tolerant capabilities[3]. In recent years, developments in PCM-based reconfigurable and tunable photonic technology have given rise to the notion of PCMs as a material platform enabling versatile, compact spaceborne photonics[4,5]. Such multifunctional photonic devices are expected to find LEO applications such as optical modulators in integrated photonic modules aboard miniaturized satellites (SmallSats / CubeSats), reconfigurable optical devices in remote sensing missions, multifunctional lab-on-a-chip astronaut health monitoring systems, and rapid environmental sensors on the NASA Orion crew transport vehicle[5]. Especially, reconfigurable photonics based on PCMs allow dynamic tuning of optical functionalities post-fabrication and thus have opened up exciting opportunities for unexpected new discoveries in space from agile manipulation of light[5]. The function and properties of traditional optical devices are set once fabricated[6].

This comment discusses the recent advances in rapidly growing PCM research and its potential to transition from a conventional terrestrial optoelectronics materials platform to a versatile spaceborne photonic materials platform (Fig.1a) for current and next-generation missions. Based on the underlying physical characteristics of PCMs, harsh space environment effect on PCMs through the Materials International Space Station Experiment-14 (MISSE-14) mission following experimental demonstrations in novel optical sub-systems, and associated space performance metrics, we argue for the technology's potential for widespread integration in space systems. Discussions on space material definitions are presented via NASA space science mission-driven examples. Due to remaining unknowns such as material long-term reliability and maximum cycling, we intend for the arguments presented herein to help direct future PCM research towards the expanding field of extraterrestrial applications, as well as to help inform and guide future funding of space technology research more optimally.

**Phase-change materials: From terrestrial to space applications**
PCMs are solid-state materials that can change between amorphous and crystalline phases as a result of thermal stimulus[7–9]. This is accompanied by a large shift in the material's electronic and optical properties (Fig.1b) and systematic trends in properties and performance of PCMs have recently been discussed[10]. The amorphous phase boasts higher electrical resistance, a lower refractive index, and less optical absorption (lower extinction coefficient), while the crystalline state shows the exact opposite. The magnitude of these changes can also be quite dramatic. Typical PCMs have a wavelength-dependent refractive index shift much greater than unity (typically ~1–2)[11] accompanied by a small increase in extinction coefficient, and resistivity shifts of 3 orders of magnitude[12]. It is also possible to exploit a near-continuum of partially-crystalline states, with properties that lie more-or-less linearly between those of the amorphous and crystalline states[4]. Importantly, all of these phase changes are nonvolatile, meaning they do not require a constant supply of energy to maintain their phase change; i.e., '*set it and forget it*' if you will. The different compositions of germanium (Ge)-antimony(Sb)-telluride(Te) alloys (Fig.1c) present the most widely utilized PCMs[13]. In GeSbTe (GST), the crystallization temperature is a function of composition (Fig.1d), with different applications suiting different GST alloys[14]. These *phase changes* occur on a very short timescale (between tens of milliseconds and a few picoseconds)[7,8], depending on the material type or composition.

The concept of applying different, non-chalcogenide PCMs to space applications is not entirely unheard of. Since the 1970s[15], gallium and paraffin materials (materials with a solid-liquid phase changes) have been proposed and tested for spacecraft thermal control in LEO or lunar orbit[16]. However, to-date, the application of all-solid-state, and particularly chalcogenide, PCMs for space applications has been largely non-existent.

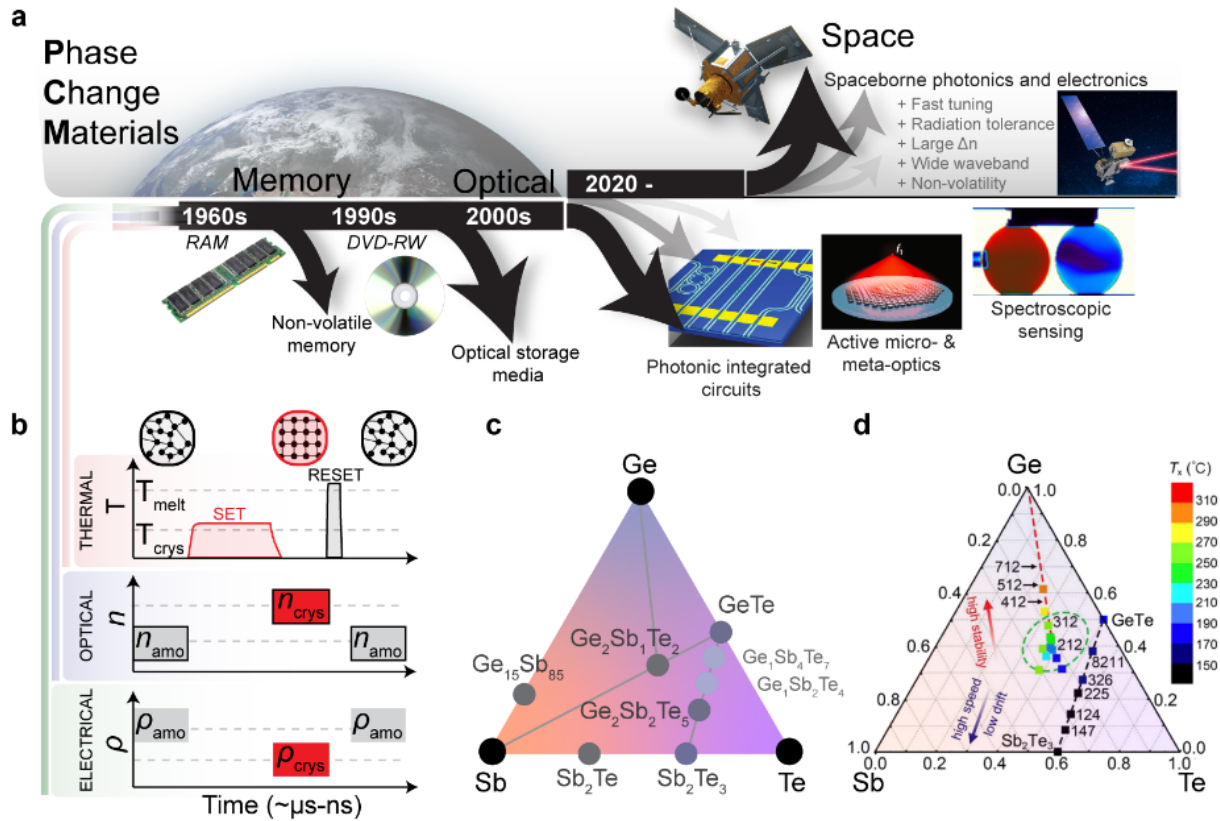

**Fig.1| Prototypical phase-change materials and their characteristics, application-driven interest of PCMs since 1960.** (a) Application-driven interest of PCMs since 1960: from terrestrial to spaceborne applications. (b) Switching the crystalline phase into the amorphous phase involves heating above $T_{melt}$ (melting point) and then rapidly quenching the PCM. Switching the amorphous phase into a crystalline phase involves heating above $T_{crys}$ (crystallization temperature) for sufficient time to crystallize the PCM. PCM devices are programmable via the "SET" state (akin to writing a logic "1") and "RESET" state (writing a logic "0"). (c) Ge-Sb-Te ternary phase diagram showing popular GeSbTe (GST)-based PCMs include $Ge_2Sb_2Te_5$, $Ge_1Sb_4Te_7$, and $Ge_1Sb_2Te_4$. (d) Crystallization temperature as a function of composition in the Ge–Sb–Te ternary phase diagram. The optimal compositional region for high-performance embedded phase change memories is highlighted in green. Adapted from Liang Sun *et al.* paper[14].

**MISSE-14 mission flown PCMs outside of the ISS and the results provide strong evidence of the applicability of spaceborne photonics using PCMs[17]**

Key advances have also been made in material survivability qualification through the *Materials International Space Station Experiment-14 (MISSE-14)* mission[18]. Although PCMs have previously been noted for their resilience to various forms of radiation[3], they had not been long-term exposed in a realistic space environment until 2021. Twenty-four samples of various PCM thin films, along with optical filters comprised of these PCMs[18], were delivered to the International Space Station (ISS) via a Northrup Grumman (NG) Cygnus spacecraft. These were then installed on the MISSE-Flight Facility (*MISSE-FF, in a LEO about 400km to 420km above the Earth's surface*) and the MISSE-14 Science Carriers (MSCs) opened to start exposure of the specimens to the space environment. The samples were flown outside of the ISS and then returned on a SpaceX CRS-24 capsule (also known as SpX-24) in January 2022 as shown in Figure 2a. The total open exposure time for the specimens was 133 days 5 hours 6 minutes in the Wake direction and 148 days 21 hours 11 minutes in the Zenith direction. The samples were returned to NASA Langley Research Center in March 2022 for post-flight characterization.

Space materials and photonics on the exterior of a spacecraft will be subjected to many environmental effects and threats that can cause degradation. In LEO these threats include visible light photon radiation, ultraviolet (UV) radiation, vacuum ultraviolet (VUV) radiation, x-rays, solar wind particle radiation (electrons, protons), cosmic rays, temperature extremes, thermal cycling, impacts from micrometeoroids and orbital debris, on-orbit contamination, hard vacuum, and atomic oxygen (AO)[19]. Such harsh environmental exposures can result in optical property degradation of PCMs and filter performance (i.e., center wavelength, bandwidth, transmittance) and filter tunability (i.e., tuning range, speed, switching cycles) threatening spacecraft performance and durability.

The PCMs were exposed in space along the two different orientations of Zenith (deep space view, away from earth and grazing AO and highest solar / most UV exposure) and Wake (general space exposure, away from the direction of ISS travel and moderate solar / UV exposure). In-space monitoring data included the ionizing radiation dose (Fig.2b), UV irradiance (Fig.2c), and temperature (Fig.2d). Surface damage/contamination photographs (Fig.2e) were included. Ionizing radiation dose and total AO fluence data were provided at the conclusion of the mission.

During the mission, the MSCs were closed 35 times (Wake) and 39 times (Zenith) for possible contamination events (i.e., docking, undocking, thruster tests, relocation, reboost, deboost, and prop purge). The estimated total surface contamination during the MISSE-14 flight was 7.96 Å (Wake) and 0.37 Å (Zenith) bonding thicknesses at +25 °C. AO, which is the most prevalent atomic species encountered in LEO, is highly reactive with plastics and some metals causing severe erosion. There is also extreme ultraviolet (UV) radiation due to the lack of an atmospheric filter. This radiation deteriorates and darkens many plastics and coatings. The vacuum in space also alters the physical properties of many materials. Impacts of meteoroids and orbiting man-made debris can damage all materials exposed in space. The combined effects of all of these environments on specimens and photonic devices can only be investigated in space. On Earth, a material can only be subjected to one environment at a time. Samples on the Zenith and Wake carriers experienced a total AO fluence of 3.07E+19 atoms/cm$^2$ and 3.96E+19 atoms/cm$^2$ exposed area, respectively. The AO fluence was determined[20] using a mass loss technique using Kapton witness samples with an exposed area of 2.742 cm$^2$. AO can induce PCM degradation and erosion by producing surface recession and mass loss which have the most severe impact on optic characteristics such as transmittance. An AO-resistant $SiO_2$ coating can be applied[21] to address this issue.

The monthly measured ionizing dose from the sensor (Teledyne uDOS001-C radiation detector shielded by 0.3 – 0.5 inch thickness aluminum) located on the Zenith MSC was 0.112 rads (with the measurement started on June 30), 1.078 rads (July), 2.132 rads (August), 1.42 rads (September), 3.096 rads (October), 2.201 rads (November), and 1.956 rads (measurement ending on December 26). All of these values are within the recorded average[22]. Even with radiation hardness considered an issue for materials in LEO, no crystallization peaks were produced on the MISSE-14 flight PCMs ($Ge_2Sb_2Te_5$, $Ge_2Sb_2Se_4Te_1$, $Sb_2S_3$) based on x-ray diffraction measurements except for two peaks on the $CaF_2$ substrate (28.227° for $2\theta_{111}$ and 58.435° for $2\theta_{222}$). Moreover, PCMs, in particular GST, are reported to be tolerant to ionizing radiation effects from the appreciable void volume present in the amorphous state as elucidated by ab initio molecular-dynamics simulations[3].

It is suspected that extreme temperature cycling of -120 °C to 120 °C[23], with sixteen cycles per day for the ISS, imparts the most impact since this can lead to degradation of data retention periods related to the thermal stability of the amorphous state of PCM. However, the actual measured temperature (using a k-type thermocouple, Nanmac A14B-1-24-K-48 model and attached to the underdeck of the deck on both the swing and mount sides of the MSC) was in a -20 °C to 50 °C range, not matching the extreme temperature ranges in the literature. The temperature is impacted by the ISS flight orientation and the Sun Beta Angle which provides insight into the temperature flux during the mission. These lower temperature fluctuations contributed to a high amorphous-phase stability of the PCM. The vacuum environment of LEO, with typical pressures less than $10^{-6}$ Torr, is similar to the base pressure of solid-state PCM film deposition and would not lead to outgassing following mass change of materials or system contamination. Along with the science data, high-resolution cameras scanned and captured photographs of the samples about once a month to detect changes as a function of time on-orbit.

Figure 2f shows the pre-launch and post-flight characterized transmission spectra results of: (1) four $Ge_2Sb_2Te_5$-based Fabry-Perot (GST-FP) bandpass filters with different center wavelengths (CWLs, 3.46 μm, 3.60 μm, 4.26 μm, and 4.7 μm) across the Mid Wavelength Infrared (MWIR), and (2) broadband dielectric mirror and $CaF_2$ substrate references. The measurements indicate a slight reduction in performance of the bandpass filters at 3.46 μm after 6 months of LEO exposure. The average peak transmission decreased 5-10% (highlighted in detail in Figure 2f, *right*), with minor broadening of the passband FWHM and a minor blue-shift of the CWL (both approximately several nanometers in scale).

Space-environment induced degradation of multi-layer optical filters and optical coatings has been previously reported[24–28]. Using these studies, which include detailed analyses of failure mechanisms, the transmission decrease and passband broadening observed in PCM-filters may be associated with small reductions in reflectance between the multi-layer thin-film interfaces, due to (1) layer interdiffusion or mechanical stresses arising from extended thermal / vacuum cycling[24,25,27,28], (2) contamination from water molecules and outgassing, which increases off-axis scattering[24,25], or (3) reactive AO[28]. The small CWL wavelength shift in space-based filters may be linked to compaction of thin-films due to large thermal cycling[24]. The degradation associated with ionizing radiation and UV solar radiation—usually predominant at the shorter wavelengths (i.e. UV) and negligible at longer (i.e. MWIR) wavelengths[27]—is discounted as an explanation for the MWIR transmittance results. However, visual inspection of one of the bandpass filters (3.46 μm CWL filter in Figure 2e) shows color variation across its surface, indicative of AO bombardment[28], but yields negligible transmittance variation in the MWIR. It can be concluded that the PCM-based filters have performance degradation in-keeping with similar 'hard coating' optical filters[24–28], that is, minimal changes in transmittance and passband FWHM, for LEO exposure. Significant changes in performance due to PCM-integration has not been observed, and filter function—closely resembling pre-launch performance—remains across all filters. We further show 'real-world' applicability by performing $CO_2$ gas sensing (Fig.2g) and multispectral thermal imaging (Fig.2h) with space exposed 3.46 μm and 4.26 μm CWL filters, respectively. For the thermal imaging demonstration, the GST-FP bandpass filter (3.46 μm CWL) was annealed at 200 °C and 400 °C inside a vacuum chamber to induce GST crystallization and continuously tunable (i.e., a-GST with 3.46 μm, $p_1$-GST with 3.72 μm, $p_{final}$-GST with 4.38 μm CWL) operation was realized. In addition, the filter, designed to have its initial CWL (amorphous GST sate) matched to a molecular (vibrational) absorption mode of $CO_2$, successfully demonstrates the gas sensing image of externally added $CO_2$ gas. Two separate filters (3.6 μm and 4.26 μm) have been used for demonstration purpose here, but clearly a single, switchable, GST-FP tunable bandpass filter of this originally design has the $CO_2$ gas detection ability even after six months in space.

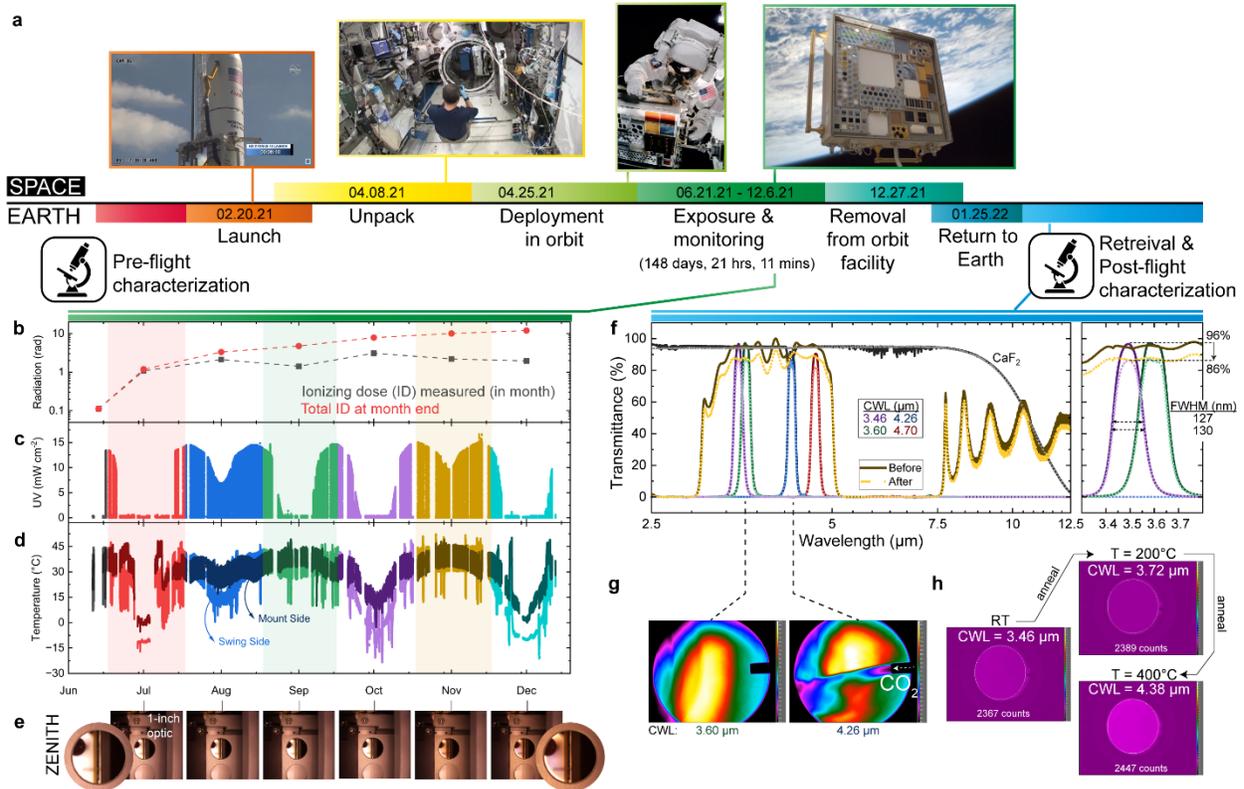

**Fig.2| Materials International Space Station Experiment-14 (MISSE-14) PCM and PCM-based Tunable MWIR Filters Experiment.** (a) Overview of MISSE-14 mission operation cycle with on-orbit images (Credit: NASA). MISSE-14 environment flight data, highlighting (b) ionizing radiation, (c) ultraviolet radiation (UV), (d) temperature, and (e) on-orbit photographic images at the 3.46 μm center wavelength (CWL) filter from June to December 2021 on Zenith. (f) Characterized transmission spectra of $Ge_2Sb_2Te_5$ (GST) Fabry-Perot (GST-FP) bandpass filters with different CWLs (3.46 μm, 3.60 μm, 4.26 μm, and 4.7 μm) before (pre-flight) and after (post-flight) the LEO exposure. A MWIR broadband dielectric mirror and $CaF_2$ substrate (optical window) is also included for reference. (g) LEO exposed GST-FP bandpass filters imaging: $CO_2$ gas sensing with GST-FP filters and (h) MWIR imaging results at a fixed 363 K blackbody spectral irradiance, as a function of varying GST-FP bandpass filter states from thermal annealing, with varying passband CWLs.

**The emerging applications of PCM-based photonics for space science missions**
The emerging PCM-enabled spaceborne applications is summarized in Figure 3a,b, whereby notable advancements in terrestrial applications can be exploited for the space sector. Broadly speaking, PCM-devices enable low-SWaP and high performance for a number of capabilities with relevance to space missions (listed alongside the capabilities in parenthesis), such as: (1) photonic integrated circuits (high speed communications and sensing), (2) LIDAR and imaging spectroscopy (spatial light modulators, beam steerers, tunable filters), (3) deep-space imaging (autofocus/real-time phase-corrective lenses, planar adaptive optics), and (4) satellite temperature management/thermal homeostasis (tunable/dynamic thermal emission control)[2]. Mission requirements such as operating waveband and modulation speed restricts PCM selection. Figure 3c,d collates the wavelength dependent change in refractive index and extinction coefficient for different PCMs[9,29,30], and Figure 3e shows the modulation (switching) speed for a reduced selection[2,5]. Operation at shorter wavelengths (VIS)—necessary for conventional optical imaging—mean most PCMs are unattractive due to their high extinction (loss), apart from niche applications such as color coatings or e-ink. In near-infrared (NIR) applications—such as astronaut health monitoring and telecommunications—SbS and GSST alloys become more attractive. Further into the IR lies regions for atmospheric gas monitoring and vehicle thermal imaging, where GST becomes more suitable. Performance parameters, including switching speed and contrast, endurance (lifetime), and power consumption can vary depending on the specific use cases[2,5] and their 'importance' further dictates the suitability of specific PCMs (Fig.3f).

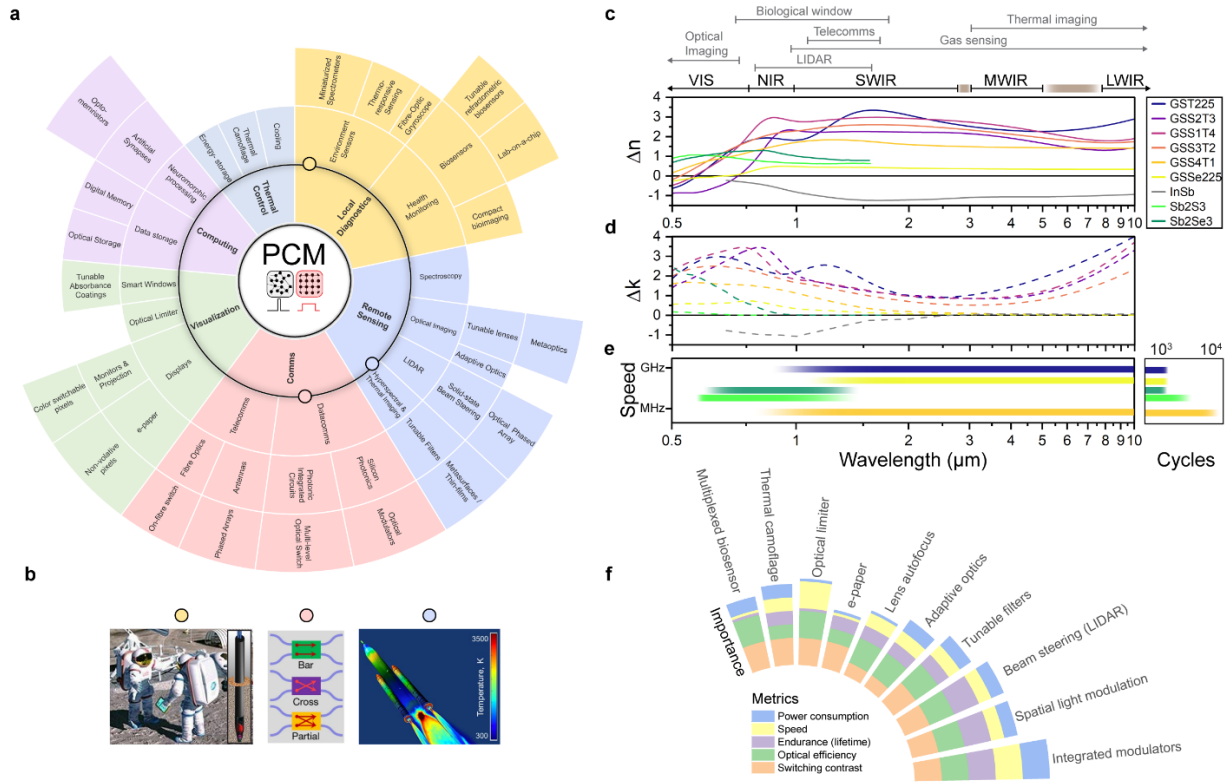

**Fig.3| PCM-based photonics for spaceborne applications:** (a) Hierarchical illustration of the applications, technologies, and sub-systems (radially outward respectively) enabled by PCM-integrated optical devices. (b) Three application areas are visually highlighted, (*left-to-right*): from miniaturized spectroscopic analysis of lunar soil [local diagnostics]; tunable photonic integrated circuit elements [data communications]; and scientifically calibrated in-flight thermal imaging of heat shields [remote sensing]. (c-e) Property comparison for different phase-change materials as a function of operating wavelength with major application-specific wavebands highlighted above, with: (c) change in refractive index ($\Delta n = n_{crys} - n_{amor}$); (d) extinction coefficient ($\Delta k = k_{crys} - k_{amor}$)[9,29,30]; (e) modulation speed and maximum switching cycle (experimental)[2,5]. (f) Visual representation of the importance of PCM performance metrics for different applications; adapted from Mikhail Y. Shalaginov *et al.* paper[2].

The first viable technology for space applications may be solid-state tunable filters. The PCM-based tunable optical filter is an 'all encompassing' acronym—technically it includes any filter design, which incorporates a PCM as the tunable constituent to effectively tune the transmission passband. Such filters have recently been shown for a variety of device architectures[4,31,32], operating across the NIR to MWIR wavebands. Thermography and imaging spectroscopy (IS) are critical measurement techniques for a variety of NASA missions including vehicle ascent and vehicle Entry, Descent, and Landing (EDL) projects. The PCM-based tunable filter offers significant advantages over the current IS state-of-the-art via the ability to remove the traditional bulky motorized filter wheel used in current IS systems and replace it with a single, low-power-consumption filter, with no moving parts. Apart from the orders-of-magnitude reduction in total SWaP, it also offers significant advantages in terms of data collection, as PCM tuning speeds allow for much higher temporal resolution than rotating filter wheel mounts. Additionally, the PCM-based tunable filter technology can support human exploration and operational missions in space through astronaut health monitoring, including detection of specific blood-based or tissue-based molecular markers for in-situ discovery and monitoring of abnormal conditions (diseases), or ex-situ analysis[33].

Beyond filters, PCM-based reconfigurable optical wavefront lenses are attractive for exoplanet imaging. Exoplanet imaging in space requires real-time wavefront corrections to mitigate the effects of thermal gradients, optical imperfections, and diffraction issues[34]. Current methods of performing these corrections involve deformable mirrors, requiring many actuators to provide the necessary authority to the mirror. Recent demonstrations suggest that PCM optics could be used to simplify the correction system and introduce a transmissive correction element versus a reflective one[35,36]. The PCM-based optical wavefront lens would be beneficial for the application since the wavefront correction system has to exhibit the control authority to compensate for very small aberrations at high speed[37]. The fast-tuning capability of PCM-based optics is by far the most important feature since exoplanet imaging requires control of high frequency spatial and temporal aberrations.

Free-space optical (FSO) communication is another application area for PCM-based photonics, where high fidelity laser beam steering is critical. Recent demonstrations in reconfigurable PCM-based optics[38–40] suggest that such devices could be utilized at both the transmission and receiving ends of a FSO system to remove the effects of vibration and thermal gradients as replacements for deformable mirrors. This would permit higher bandwidth optical communications through space. Additionally, PCM beam steerers with broad angular ranges can be used to maintain optical links across satellite constellations without the need to maneuver satellites into a direct line-of-sight configuration, reducing fuel usage and coordination complexity across the constellation. Moreover, the small volume of the PCM-based photonics is important for operation in CubeSat platforms in space.

**Application 1 - Tunable bandpass filter for remote sensing: Spaceborne LIDAR, IR imaging**
Because of their rapid tuning speeds and broadband transparency in the Short Wavelength Infrared (SWIR), to Long Wavelength Infrared (LWIR), PCM-based optics are a strong contender to make an impact on space-based LIDAR and IR imaging telescopes. The reduced SWaP of PCM-based metasurfaces compared to traditional optical components like filter wheels and spectral splitters allows for a significantly reduced SWaP that is amenable to small satellite form factors.

As an example, consider the Stratospheric Aerosol and Gas Experiment (SAGE) missions, a series of remote sensing instruments that measure key atmospheric constituents, with specific aerosols and molecules of interest[41]. Unlike previous SAGE instruments, which were mounted on free-flying spacecraft of different sizes, SAGE IV is 6U CubeSat platform concept for observing key atmospheric constituents at the same quality as previous SAGE instruments that furnishes both benefits and challenges. The CubeSat platform (Fig 4a) allows high vertical resolution measurements with semi-global coverage of key trace gas constituents in the stratosphere and upper troposphere. However, the 6U form factor requires a significant size, weight, power, and cost (SWaP) reduction, on the order of ~1/10$^{th}$. SAGE uses a filter wheel (Fig .4c) to switch between different chemical absorption channels such as $H_2O$, $N_2O$, $CO_2$ and $CH_4$[41]. This requires tuning to ~6 discrete wavelengths from 386 nm to 1020 nm.

PCM-based filters may facilitate a SAGE IV scenario with significant SWaP reduction in 6U CubeSat architectures by empowering a dynamically tunable, all-solid-state solution without moving parts[4,32,33] (Fig.4f). For the SAGE IV scenario, arguably one of the strictest science requirements to which they can be applied – mainly, (1) materials science challenges of identifying a PCM that can operate efficiently across visible and near infrared wavelengths and (2) good cyclability for electronic PCMs levels—they must meet or exceed these specifications. Careful design of the PCM metasurface filter—for example, via deep neural networks[31]—along with proper choice of the PCM material can likely satisfy the requirements for tuning range and transmission passband width. The switching requirement seems likely to be trivial for PCM devices, as they are currently used in PRAM electronics. However, switching stability and failure analysis for photonic PCMs has yet to be fully characterized, and is a key research element required to transition PCMs from the lab and into scientific instruments. SAGE switches between filters at a rate that translates to $\sim 9.5 \times 10^6$ switching cycles per year; the science observation events are a maximum length of 6 minutes, and the filter (specification of high-speed motorized filter wheels, ~1 ms per filter) will be switching five times per second during the event. LEOs are approximately 90 minutes long, and one science observation event will take place per orbit. Approximately $10^7$ cycles are performed per year on orbit based on this estimation. There is potential interest in using the SAGE IV chassis for future missions, optimizing its IR (1~5 μm) measurement capability. This requires tuning to ~8 discrete wavelengths from 1.4 - 4.1 μm for enhanced science measurements (i.e., ×10 better $H_2O$ measures) and for extended species detection (i.e., $CH_4$, $CO_2$, $N_2O$, and CO) [42]. Moreover, the reduced SWaP benefits from using PCM-based filters as a replacement for filter wheels reduce volume requirements, thereby allowing the addition of a spectrometer. The SAGE IV IR is applicable for Mars ($CO_2$, $N_2$, $H_2O$, NO) and / or Venus ($CO_2$, $N_2$, $H_2O$, CO) atmospheric monitoring.

**Application 2 - Reconfigurable planar optics for wavefront correction and beam steering: Spaceborne LIDAR, Free-Space Optical Communications**

A second area of interest is within applications that require adaptive optics for wavefront correction or beam steering. The use of MEMS-based micromirror arrays and / or fine-steering mirrors (FSMs) is common in various imaging LIDAR and FSO communications systems (Fig.4b) in order to capture high-spatial-resolution data within a given field of regard, correct for wavefront aberrations caused by scatter/turbulence, or maintain optical links—both between satellites and ground links (Fig.4d)—without having to maneuver satellite constellations to maintain direct line-of-sight. Recent work in the development of FSO communications has concentrated on adapting this technology for use in SmallSat and CubeSat platforms since FSO communication systems can potentially increase signal to noise ratios significantly within the communication channel[43]. However, the actuator response times and pointing requirements for these applications can be strict. PCM-based tunable metasurfaces such as beam steerers and tunable metalenses have recently been demonstrated as proof-of-concept devices, and have significant potential to address both of these applications[35,44]. Due to their large refractive index contrast, PCM-based beam steerers and highly pixelated "meta-correctors" can achieve a broad range of phase coverage, translating to large angular scanning and phase-correction ranges (Fig.4e). Although these devices have only recently been demonstrated in the lab, targeted research into high-resolution control of intermediate PCM phases and highly pixelated metasurfaces could soon play a transformative role in space-based adaptive optics such as the aforementioned applications.

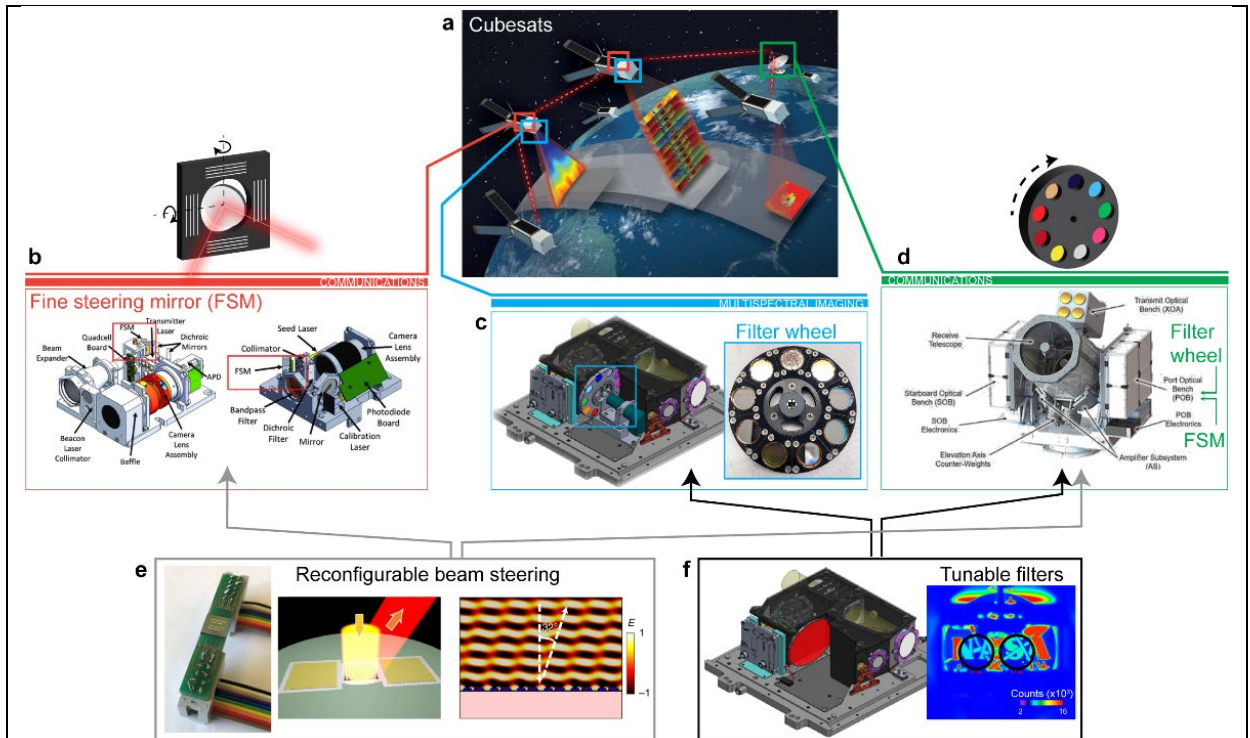

**Fig.4| Spaceborne application examples: PCM-integrated active devices for technology subsystems in CubeSats.** (a) A constellation of CubeSats providing time-resolved spectroscopic imaging. Image credit: MIT/LL, NASA[45]. (b) CubeSat Laser Infrared CrosslinK (CLICK) payloads for space-to-ground and crosslink communications. The fine steering mirror (FSM) steers the beams (976 nm Beacon laser / 1550 nm transmitter laser) to dichroic mirrors in the respective optical systems[46]. (c) SAGE-IV (Stratospheric Aerosol and Gas Experiment-IV) pathfinder multispectral LIDAR instrument, with filter wheel consisting of eight bandpass filters and one opaque element. Each designed to observe a target species such as aerosols and ozone[41]. (d) NASA's LCOT (Low-Cost Optical Terminal) FSOS (Free-Space Optical Subsystem), a ground terminal for space optical communications[47]. This terminal consists of two sets of filter wheels and a FSM. (e) Reconfigurable planar optics for free-space optical communications: PCM-based beam steerers and highly pixelated "meta-correctors" can achieve a broad range of optical phase coverage, translating to large angular scanning and phase-correction ranges[38] (e, *left*). Packaged device consisting of the reconfigurable PCM-metasurface (*middle, right*) which deflects an incoming IR beam with output angle dependent on index of the PCM. (f) Tunable bandpass filter for spaceborne LIDAR: (f, *left*) The SAGE filter wheel may be replaced with a single element PCM tunable filter used to switch between different CWLs (i.e., different chemical absorption channels). Through an optical or electrical stimuli on the filter active area, PCM crystallinity is modified, and the resultant transmission response ($\lambda_N$) is spectrally shifted as demonstrated (*right*) by $CO_2$ gas sensing with GST-Fabry Perot tunable bandpass filters[48].

**Outlook: challenges and opportunities toward space applications**

The definition and qualification of space materials is challenging. Understanding the performance of materials under extreme conditions sets bounds to where and how they can be used and informs the range of system engineering that may be needed to enable their use in applications. Similarly, space missions come with extremely challenging and wholly inflexible performance metrics to ensure mission success, and understanding the capabilities of devices comprised of such materials with respect to inflexible mission metrics is equally important.

The most mature PCM demonstrations have been carried out as proof-of-concept or low-fidelity breadboard prototypes in a controlled lab environment, placing PCM devices at TRL 3-4 in the highest cases. In order to cross the so-called "Valley of Death" to TRL 9 and truly make an impact on photonic applications in space, PCMs must address a number of hurdles, spanning fundamental science to (sub-)system implementation.

- **TRL 4-5:** Photonic components deployed in space require additional performance evaluations for their ability to withstand the harsh space environment as a complete system. PCMs are a materials platform for a sub-system / technology. This can enable a particular part and / or small 'active' part of a photonic application like photonic integrated circuits for beam steering applications in CubeSat platforms. Pragmatically, this means that it is predicted that PCMs will find use in space qualified materials, optics, housing, and platforms that can withstand both the launch and space environments.

- **TRL 5-6:** The cycle lifetime or endurance of PCMs in photonic devices remains to be validated or improved[49]. In electronic memory, PCM endurance and failure mechanisms have been extensively characterized and have indicated an endurance value of over $2 \times 10^{12}$ cycles. However, these studies have been solely focused on electronic PCMs (and related properties). There is still significant effort required to fully validate optical PCM longevity (e.g., GeSbTe and GeSbSeTe), the applications for which can require similar endurance values as electronic applications depending on the duration of the space mission. This limitation is primarily a material issue, as the fundamental physics related to optical PCM switching longevity (and how they are affected by the space environment) are not yet fully understood. Reproducible PCM optical device properties are important throughout the mission period and need to be evaluated as part of the endurance. Understanding of failure modes specific to photonic devices, the fundamental thermal transport properties of PCMs[50] (to control the repeatability of the PCM switching), and implementation of the presently-used design rules to boost device lifetime[51] will be imperative to facilitating adoption of PCMs in practical applications, and is arguably the greatest challenge currently faced by the technology. Despite these material issues, system-level considerations like tight control of the electronic pulse scheme and thermal design of the PCM structure as a whole are also paramount. Minimizing the switching voltage and current of PCMs is another important direction for space applications in lean power budgets.

- **TRL 6-7:** For applications intending to perform imaging in the MWIR/LWIR range, design work is required to ensure that the thermal emission from the PCM – incurred when the PCM is heated to switch states – does not interfere with the desired signal. This is a not only a systems consideration, but also plays into the device architecture to ensure rapid cooling/quenching of the PCM, ideally on timescales faster than the desired data acquisition rate of the full system (e.g., the camera frame rate).

- **TRL 7-9:** To successfully move from prototype to true subsystem, proper electrical integration of the PCM device must be achieved. While breadboard demonstrations may use hand-soldered contacts and bulk leads and power supplies, full integration requires contacting the device in a package similar to traditional microelectronics, complete with a power supply consistent with the volume and power constraints of an intended application. In layman's terms, PCM devices need to go from bulky lab demonstrations to looking like something you might find inside of your cell phone or laptop. For certain applications, such as aberration correction and beam steering, this will also require significant pixelation of the device on the order of tens-of-thousands of pixels or more. While the present state of that particular PCM implementation sits at a much lower TRL (~ TRL 2), it must be considered nonetheless, and will likely require the involvement of proper foundry partners to achieve – as well as a better understanding of PCM thermal transport to achieve thermal isolation and minimize pixel crosstalk.

- Finally, although not required to reach higher TRLs, another key materials science challenge is identifying a PCM that can operate efficiently across visible and near infrared wavelengths. This is important for a number of spectroscopic applications, particularly aerosol remote sensing. Optical loss (non-negligible extinction) is the bane of optical PCMs. Hence, another material challenge is development of a new class of PCMs where the phase transition only triggers refractive index (real-part) modulation with a minimal loss penalty related through the Kramers–Kronig relations[9]. This would be a game changer for optical engineers that would open up numerous applications.

The space sector has witnessed tremendous growth within the past decade—not only from government agencies but also entrants from the private sector. This industry, along with societal interest in space exploration, is expected to grow even faster during this decade. The introduction of PCM technology and associated optical devices will help to accelerate the adoption of new architectures for reduced SWaP-C platforms in space. Even though there are currently no clear answers for *"What is a space material?"* and *"Who has the responsibility for a space evaluation?"*, we foresee that the arguments put forth in this comment, along with the MISSE-14 sample evaluation and data sharing within the science community, will significantly expedite PCM integration into spaceborne platforms and open emerging applications.

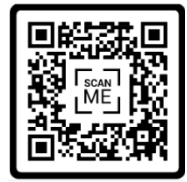

Webpage (https://spaceborne-pcms.github.io) is generated as Supporting Information for the npj Microgravity readers. Scan the QR code for "The Maps", the locations of space material evaluation facilities (academic research laboratories, space agencies, and private companies) including simulated space environments across the world.

**Acknowledgments**

The authors are grateful to Mr. Stephen Borg at NASA Langley Research Center / LaRC for his helpful technical discussions.

**Author contributions**

All authors contributed to writing the paper.

**Competing financial interests**

The authors declare no competing financial interests.

**References**

1. Space: Investment shifts from GEO to LEO and now beyond | McKinsey.

   https://www.mckinsey.com/industries/aerospace-and-defense/our-insights/space-investment-shifts-from-geo-to-

   leo-and-now-beyond.

2. Shalaginov, M. Y. *et al.* Design for quality: reconfigurable flat optics based on active metasurfaces.

   *Nanophotonics* **9**, 3505–3534 (2020).

3. Konstantinou, K., Lee, T. H., Mocanu, F. C. & Elliott, S. R. Origin of radiation tolerance in amorphous Ge2Sb2Te5 phase-change random-access memory material. *Proceedings of the National Academy of Sciences* **115**, 5353–5358 (2018).

4. Julian, M., Williams, C., Borg, S., Bartram, S. & Hyun Jung, K. Reversible optical tuning of GeSbTe phase-change metasurface spectral filters for mid-wave infrared imaging. *Optica* **7**, 746–754 (2020).

5. Gu, T., Kim, H. J., Rivero-Baleine, C. & Hu, J. Reconfigurable metasurfaces towards commercial success. *Nat. Photon.* **17**, 48–58 (2023).

6. Wright, G. S. *et al.* The Mid-Infrared Instrument for the James Webb Space Telescope, II: Design and Build. *PASP* **127**, 595 (2015).

7. Wang, Z. *et al.* Resistive switching materials for information processing. *Nat Rev Mater* **5**, 173–195 (2020).

8. Aryana, K. *et al.* Interface controlled thermal resistances of ultra-thin chalcogenide-based phase change memory devices. *Nat Commun* **12**, 774 (2021).

9. Delaney, M., Zeimpekis, I., Lawson, D., Hewak, D. W. & Muskens, O. L. A New Family of Ultralow Loss Reversible Phase-Change Materials for Photonic Integrated Circuits: Sb2S3 and Sb2Se3. *Advanced Functional Materials* **30**, 2002447 (2020).

10. Persch, C. *et al.* The potential of chemical bonding to design crystallization and vitrification kinetics. *Nat Commun* **12**, 4978 (2021).

11. Zhang, Y. *et al.* Broadband transparent optical phase change materials for high-performance nonvolatile photonics. *Nat Commun* **10**, 4279 (2019).

12. Zhang, W., Mazzarello, R., Wuttig, M. & Ma, E. Designing crystallization in phase-change materials for universal memory and neuro-inspired computing. *Nat Rev Mater* **4**, 150–168 (2019).

13. Wuttig, M. & Yamada, N. Phase-change materials for rewriteable data storage. *Nature Mater* **6**, 824–832 (2007).

14. Ab initio molecular dynamics and materials design for embedded phase-change memory | npj Computational Materials. https://www.nature.com/articles/s41524-021-00496-7.

15. Kelliher, W. & Young, P. Investigation of phase-change coatings for variable thermal control of spacecraft. (1972).

16. Howes, A., Nolen, J. R., Caldwell, J. D. & Valentine, J. Near-Unity and Narrowband Thermal Emissivity in Balanced Dielectric Metasurfaces. *Advanced Optical Materials* **8**, 1901470 (2020).
13

17. Kim, H. J. Versatile Spaceborne Photonics with Chalcogenide Phase-Change Materials Supporting Webpage. https://spaceborne-pcms.github.io/.

18. Boucher, M. NASA Space Station On-Orbit Status 14 May, 2021. *SpaceRef* https://spaceref.com/space-stations/nasa-space-station-on-orbit-status-14-may-2021/ (2021).

19. Ciminelli, C., Dell'Olio, F. & Armenise, M. N. *Photonics in Space: Advanced Photonic Devices and Systems*. (WORLD SCIENTIFIC, 2016). doi:10.1142/9817.

20. De Groh, K. K. *et al.* MISSE 2 PEACE Polymers Atomic Oxygen Erosion Experiment on the International Space Station. *High Performance Polymers* **20**, 388–409 (2008).

21. Reddy, M. R. Effect of low earth orbit atomic oxygen on spacecraft materials. *JOURNAL OF MATERIALS SCIENCE* **30**, 281–307 (1995).

22. Dachev, T. P. *et al.* Overview of the ISS Radiation Environment Observed during the ESA EXPOSE-R2 Mission in 2014–2016. *Space Weather* **15**, 1475–1489 (2017).

23. Best, F. A., Adler, D. P., Pettersen, C., Revercomb, H. E. & Perepezko, J. H. On-orbit absolute temperature calibration using multiple phase change materials: overview of recent technology advancements. in *Multispectral, Hyperspectral, and Ultraspectral Remote Sensing Technology, Techniques, and Applications III* vol. 7857 112–121 (SPIE, 2010).

24. Blue, M. D. Susceptibility of electro-optic components to degradation in a space environment. *Radiation Measurements* **26**, 807–815 (1996).

25. Blue, M. D. & Roberts, D. W. Effects of space exposure on optical filters. *Appl. Opt., AO* **31**, 5299–5304 (1992).

26. Panetta, C. J., Fuqua, P. D., Barrie, J. D., Chu, C. T. & Alaan, D. R. Lessons Learned from Optics Flown on the Materials International Space Station Experiment. in *Optical Interference Coatings (2013), paper MA.4* MA.4 (Optica Publishing Group, 2013). doi:10.1364/OIC.2013.MA.4.

27. Wernham, D. & Piegari, A. Chapter 22 - Optical coatings in the space environment☆. in *Optical Thin Films and Coatings (Second Edition)* (eds. Piegari, A. & Flory, F.) 789–811 (Woodhead Publishing, 2018). doi:10.1016/B978-0-08-102073-9.00022-9.

28. Hawkins, G. J. *Spectral Characterisation of Infrared Optical Materials and Filters*. (University of Reading, 1998).

29. Kim, H. J., Sohn, J., Hong, N., Williams, C. & Humphreys, W. PCM-net: a refractive index database of chalcogenide phase change materials for tunable nanophotonic device modelling. *J. Phys. Photonics* **3**, 024008 (2021).

30. Michel, A.-K. U. *et al.* Using Low-Loss Phase-Change Materials for Mid-Infrared Antenna Resonance Tuning. *Nano Lett.* **13**, 3470–3475 (2013).

31. An, S. *et al.* Deep neural network enabled active metasurface embedded design. *Nanophotonics* **11**, 4149–4158 (2022).

32. Sreekanth, K. V. *et al.* Dynamic Color Generation with Electrically Tunable Thin Film Optical Coatings. *Nano Lett.* **21**, 10070–10075 (2021).

33. Yoon, J. *et al.* A clinically translatable hyperspectral endoscopy (HySE) system for imaging the gastrointestinal tract. *Nat Commun* **10**, 1902 (2019).

34. Cahoy, K. L., Marinan, A. D., Novak, B., Kerr, C. & Webber, M. Wavefront control in space with MEMS deformable mirrors. in *MEMS Adaptive Optics VII* vol. 8617 42–57 (SPIE, 2013).

35. Shalaginov, M. Y. *et al.* Reconfigurable all-dielectric metalens with diffraction-limited performance. *Nat Commun* **12**, 1225 (2021).

36. Ding, F., Yang, Y. & Bozhevolnyi, S. I. Dynamic Metasurfaces Using Phase-Change Chalcogenides. *Advanced Optical Materials* **7**, 1801709 (2019).

37. Clampin, M., Rieke, M., Rieke, G. & Krist, J. ESA Science & Technology - Coronagraphic Detection of Exosolar Planets with the James Webb Space Telescope. (2019).

38. Zhang, Y. *et al.* Electrically reconfigurable non-volatile metasurface using low-loss optical phase-change material. *Nat. Nanotechnol.* **16**, 661–666 (2021).

39. Zou, L., Cryan, M. & Klemm, M. Phase change material based tunable reflectarray for free-space optical inter/intra chip interconnects. *Opt. Express, OE* **22**, 24142–24148 (2014).

40. Kim, J. *et al.* Tunable metasurfaces towards versatile metalenses and metaholograms: a review. *AP* **4**, 024001 (2022).

41. Leckey, J. P., Damadeo, R. & Hill, C. A. Stratospheric Aerosol and Gas Experiment (SAGE) from SAGE III on the ISS to a Free Flying SAGE IV Cubesat. *Remote Sensing* **13**, 4664 (2021).

42. Kim, H. J. *et al.* P-ACTIVE Project Report. (2023).

43. Dolinar, S., Moision, B. & Erkmen, B. Fundamentals of free-space optical communications. *Keck Institute for Space Studies (KISS) Workshop on Quantum Communication, Sensing and Measurement in Space\* (2012).

44. de Galarreta, C. R. *et al.* Nonvolatile Reconfigurable Phase-Change Metadevices for Beam Steering in the Near Infrared. *Advanced Functional Materials* **28**, 1704993 (2018).

45. Mission Overview | TROPICS. https://tropics.ll.mit.edu/CMS/tropics/.

46. Serra, P. C. *et al.* CubeSat laser infrared crosslinK mission status. in *International Conference on Space Optics — ICSO 2020* vol. 11852 1443–1456 (SPIE, 2021).

47. Thompson, P. L. *et al.* NASA's LCOT (low-cost optical terminal) FSOS (free-space optical subsystem): concept, design, build, and test. in *Free-Space Laser Communications XXXV* (eds. Hemmati, H. & Robinson, B. S.) 34 (SPIE, 2023). doi:10.1117/12.2653355.

48. Williams, C., Hong, N., Julian, M., Borg, S. & Kim, H. J. Tunable mid-wave infrared Fabry-Perot bandpass filters using phase-change GeSbTe. *Opt. Express, OE* **28**, 10583–10594 (2020).

49. Electrical Programmable Low-loss high cyclable Nonvolatile Photonic Random-Access Memory. https://www.researchsquare.com (2022) doi:10.21203/rs.3.rs-1527814/v1.

50. Aryana, K. *et al.* Toward accurate thermal modeling of phase change material based photonic devices. Preprint at https://doi.org/10.48550/arXiv.2305.14145 (2023).

51. Popescu, C.-C. *et al.* Learning from failure: boosting cycling endurance of optical phase change materials. in *Photonic and Phononic Properties of Engineered Nanostructures XIII* vol. 12431 39–47 (SPIE, 2023).